\documentclass{appolb}
\usepackage{graphicx}

\graphicspath{{plot/}}

\begin{document}

\title{NUCLEAR EFFECTS IN PROTON DECAY\footnote{Presented at the Zakopane Conference on Nuclear Physics, 
September 1--7, 2008.}}
\author{Dorota Stefan
\address{The Henryk Niewodniczanski Institute of Nuclear Physics,\\ Polish Academy of Science, Krak\'ow, Poland}
\and
Artur M. Ankowski
\address{Institute of Theoretical Physics, University of Wroc{\l}aw,\\ Wroc{\l}aw, Poland}
}

\maketitle

\begin{abstract}
An experimental observation of proton decay would be a spectacular proof of Grand Unification. Currently, the best constraint on the proton lifetime for the $p\to e^{+}\pi^{0}$ decay channel, coming from the Super-Kamiokande experiment, reaches $8\times 10^{33}$ years. To improve the measurement, much bigger detectors should be constructed. Moreover, a~better description of the bound-nucleon states and of the propagation of the proton-decay products through nuclear matter have to be developed. In this article special attention is paid to the argon nucleus because a liquid argon detector is a~promising candidate for the future large apparatus.
\end{abstract}

\PACS{13.30.Ce, 14.20.Dh, 24.10.Lx, 27.40.+z}

\section{Introduction}
An experimental detection of proton decay would be a milestone in particle physics, clarifying our understanding of the past and future evolution of the universe. Proton decay is predicted by Grand Unification Theories (GUTs), in which electromagnetic, weak and strong interactions merge into a single unified interaction. For a recent review of GUTs see Ref.~\cite{GUTS}.

For various decay channels, the best limits on the proton lifetime were set by the Super-Kamiokande experiment. Particulary interesting ones are
\begin{eqnarray*}
\tau(p \to e^+\pi^{0}) &>& 1.6\times 10^{33}\;\mbox{years}\quad(8.0\times 10^{33}\;\mbox{years}),\\
\tau(p \to \bar{\nu}K^+) &>& 2.3\times 10^{33}\;\mbox{years},
\end{eqnarray*}
see Ref.~\cite{SK}. The constraint in parenthesis is given in the recent thesis~\cite{Thesis}.

Liquid argon (LAr) detectors developed till now belong to the family of time projection chambers. Such detector, filled with 600 tons of LAr, is applied in the ICARUS experiment~\cite{Amerio}, which will start taking data in a few months from now. A set of cosmic-ray events detected during tests of the detector proved that its energy and position resolutions are excellent. Simulations of several proton decay channels in the LAr detector showed that some of them could be detected almost background-free, with efficiency close to $100\%$~\cite{Bueno}. It should be stressed that one of these channels---the decay into $K^+$ and $\bar\nu$---is very interesting from the theoretical point of view because it is favored by supersymmetric models. Incidentally, the Super-Kamiokande detector has much worse efficiency for this channel, namely $<9\%$~\cite{SK}.

In recent years, there is much interest in building a massive detector, mainly to search for proton decays. The oldest idea, coming from Japan, is
to construct the Hyper-Kamiokande detector with 550 kton of water. A~similar project of a water Cherenkov apparatus in the USA is called UNO. Finally, there is a European LAGUNA project~\cite{Autiero}, in which three different detectors (based on water, scintillator, and LAr technology) are considered. The one filled with 100 ktons of LAr has appealing features for proton decay searches because it would allow verification of several theoretical models within 10 years of data-taking. In particular, for the $p \rightarrow \overline{\nu}K^+$ mode, all the range predicted by theorists would be covered.

\section{Description of the approach}

Complete simulation of proton decay in a~nucleus have to include an accurate description of bound nucleons and rescattering of the decay products. As far as nucleon states are concerned, commonly applied approach is the local Fermi gas model. In principle, this model takes into account realistic density profile of the nucleus, approximating it with zones of a~constant density. In each zone, binding energy is constant whereas nucleon momenta range from zero to the local value of the Fermi momentum. The local Fermi gas model is quite complex, however, it completely ignores shell structure of the nucleus and correlations between nucleons.

The impulse approximation approach can be free of such simplifications: one assumes only that the nucleus can be described as a~collection of independent nucleons, i.e. neglecting collective behavior of nucleus constituents. Then, full characteristics of the nucleus is the distribution of momenta and energies of the nucleons, called spectral function. Thanks to accounting for the nuclear shell structure and nucleon short-range correlations, realistic spectral functions provide more accurate description of nuclear effects than the local Fermi gas model, compare figures in Refs.~\cite{ref:Bus&Leitner&Mosel&Alvares-Ruso} and~\cite{ref:Benhar&Farina&Nakamura}. Therefore it is interesting how the differences between these approaches influence simulations of proton decay~\cite{ref:Rubbia}.

In the case of argon, only approximated spectral function was obtained yet~\cite{Ankowski}. However, the analogically calculated spectral function of calcium showed very good accuracy of modeling nuclear effects in electron scattering.

During the motion through the nucleus, the products of proton decay can rescatter, producing intranuclear cascades. To deal with their simulation, we use the Bertini cascade~\cite{Cascade} implemented in the GEANT4 code~\cite{Geant4}, which works well for particles with energy below 1~GeV. It takes into account a~variety of interactions that nucleons, pions, and kaons can undergo, and uses the most recent collision cross sections.


\begin{figure}[thb]
    \begin{minipage}[l]{0.49\textwidth}
        \centering
        \includegraphics[width=0.97\textwidth]{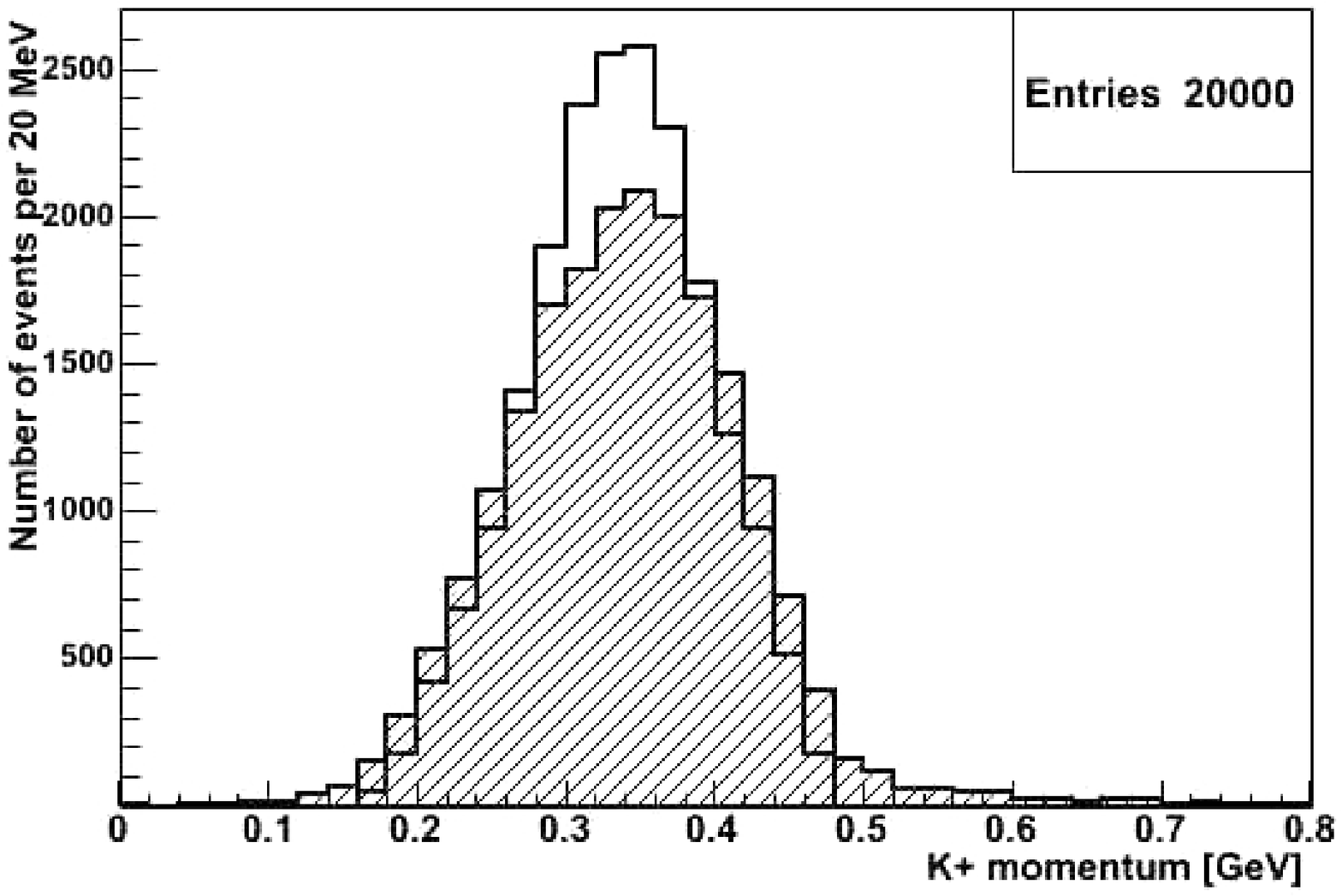}%
    \end{minipage}
    \begin{minipage}[r]{0.49\textwidth}
        \centering
        \includegraphics[width=0.97\textwidth]{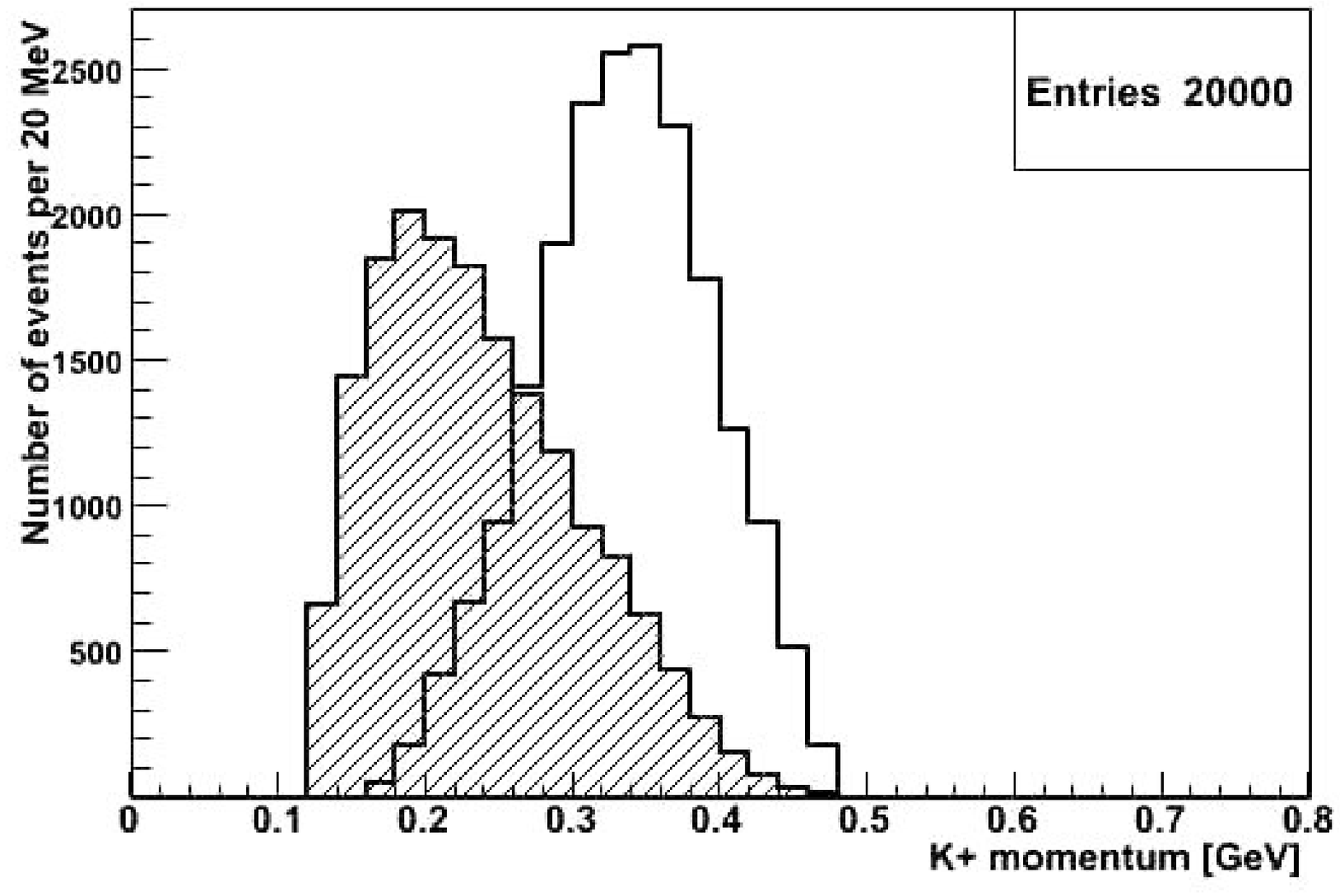}%
    \end{minipage}
\caption{Momentum distribution of kaons produced in the $p \to \bar{\nu}K^+$ decay inside the argon nucleus predicted by different approaches. Left panel: Calculations for the spectral function of argon (hatched) compared to the local Fermi gas model from GEANT4 without the intranuclear cascade (plain histogram) Right panel: GEANT4 with (hatched) and without (plain histogram) the cascade.}
\label{fig:distrib}
\end{figure}

\section{Discussion of the results}
We concentrate here on the decay of a~proton bound in the argon nucleus into $K^{+}$ and $\bar{\nu}$. The signature of such decay would be a detection of the produced kaon. The left panel of Fig.~\ref{fig:distrib} shows the laboratory momentum distribution of 20000 events generated using the local Fermi gas model and the spectral function; effect of the intranuclear cascade is not included. The considered process is a~two-body decay, so the width of the momentum distribution comes only from the Fermi motion. In the case of the spectral function, short-range correlations generate high-momentum nucleons absent in the Fermi gas model, therefore broadening is more pronounced. The peak is lower by $\sim$20\% and the allowed range of momentum is more extended.

Right panel of Fig.~\ref{fig:distrib} shows importance of the intranuclear cascade in the simulations. One can see that the cascade changes the shape of the distribution and significantly shifts its peak to lower values. The presented results concern the local Fermi gas model. Due to Pauli blocking, reinteractions can only decelerate kaons. It is not the case for the spectral function where collisions with high-momentum nucleons may also accelerate $K^+$'s.


\section{Summary}
The search for proton decay is one of the most important subjects of particle physics nowadays. To improve the upper bound on the proton lifetime, it is essential to increase the sensitivity of the measurements at least by a factor of 10. It will not be possible without an accurate description of nuclear effects. Our simulations showed that using the spectral function instead of the local Fermi gas model leads to a spreading of the momentum distributions of the decay products, lowering its peak by $\sim$20\% and enhancing the tails. It is not yet clear whether such sizable difference will be still present when the intranuclear cascade will be taken into account.

\section*{Acknowledgments}
We wish to thank Agnieszka Zalewska and Jan Sobczyk for helpful discussions. The authors were supported by MNiSW under the grants nos. 1P03B 041 30 (DS), 3735/H03/2006/31 and 3951/B/H03/2007/33 (AMA).

\end{document}